\documentclass[12pt]{article}                   
\usepackage{amssymb}
\usepackage{amsmath}
\usepackage{verbatim}
\usepackage{graphicx} 
\usepackage{cite}
\usepackage{geometry}             
\usepackage[hyperindex=true,
          pdfstartview=FitH,
          bookmarksnumbered=true,
          bookmarksopen=true,
          citecolor=blue,
          linkcolor=blue,
          colorlinks=true,
          pdfborder=001,
          unicode]{hyperref}

\newcommand{\rd}[1]{\mathrm{d}#1}
\newcommand{\pfrac}[2]{\left(\frac{\partial #1}{\partial #2}\right)}

\newcommand{\pfn}[3]{\left(\frac{\partial^{#3} #1}{\partial #2^{#3}}\right)}

\allowdisplaybreaks

\newcommand{\blue}[1]{{#1}}

\begin{document}

\title{Restricted phase space thermodynamics for \\
AdS black holes via holography}
\author{Zeyuan Gao and Liu Zhao\thanks{Correspondence author.}\\
School of Physics, Nankai University, Tianjin 300071, China
\\
{\em email}: \href{mailto:2120190129@mail.nankai.edu.cn}
{2120190129@mail.nankai.edu.cn}
and
\href{mailto:lzhao@nankai.edu.cn}{lzhao@nankai.edu.cn}
}

\date{}

\maketitle

\begin{abstract}
A new formalism for thermodynamics of AdS black holes called the {\em restricted phase 
space thermodynamics} (RPST) is proposed. The construction is based on top of Visser's 
holographic thermodynamics, but with the AdS radius fixed as a constant.  Thus 
the RPST is free of the $(P,V)$ variables but inherits the central charge and 
chemical potential as a new pair of conjugate thermodynamic variables.
\blue{In this formalism, the Euler relation and the Gibbs-Duhem equation hold 
simultaneously with the first law of black hole thermodynamics, 
which guarantee the appropriate homogeneous behaviors for the black hole mass 
and the intensive variables.} The formalism is checked in detail in the 
example case of 4-dimensional RN-AdS black hole in Einstein-Maxwell
theory, in which some interesting thermodynamic behaviors are revealed.

\end{abstract}

\section{Introduction\label{Intro}}

Black hole thermodynamics is a subject which is under intensive study for about half 
a century. Roughly speaking, the study of black hole thermodynamics can be subdivided into
two major stages. The first and the initial stage is pioneered by 
Bekenstein \cite{Bekenstein,Bekenstein2}, Bardeen, Carter and Hawking 
\cite{Bardeen,Hawking} {\em et al}. 
The formalism established in this stage may be referred to as the {\em traditional black 
hole thermodynamics} (TBHT), of which the most important accomplishment is the 
four fundamental laws of black hole physics, which hold for 
most known black hole solutions in Einstein
as well as higher curvature gravities. In this context, the black hole mass is interpreted as
the internal energy, while the surface gravity and the area of the 
event horizon are respectively proportional to the temperature and the entropy.
However, the nature of microscopic degrees of freedom which contribute 
the entropy of the black holes remains mysterious. 
\blue{According to the holographic principle \cite{Hooft,Susskind},
or better known as the AdS/CFT duality \cite{Maldacena},}
an AdS black hole state in the bulk is equivalent 
to a thermal state of the dual field theory, for which the microscopic degrees 
of freedom can be more easily understood. A parallel line of research on black 
hole thermodynamics is proposed by Wald \cite{Wald}, which recovers the thermodynamic
relations for black holes starting from various symmetries of the corresponding gravity 
models and interprets thermodynamic quantities of the black holes as Noether charges. 
Another important development in this stage is the discovery of 
Hawking-Page phase transition \cite{Hawking2} for AdS black holes. 
This is a phase transition between a black hole state and a pure thermal gas state, 
which is characterized by a vanishing Gibbs free energy.

The second major stage is the so-called {\em extended phase space thermodynamics} (EPST)
\cite{Kastor,Dolan,Dolan2,Dolan3,Kubiznak,Cai,Kubiznak2}, 
also known as {\em black hole chemistry} \cite{Kubiznak2}. 
This formalism is initiated by 
Kastor, Ray and Traschen \cite{Kastor}, and subsequently developed by a great number of 
researchers. To name a few of them, we refer to \cite{Dolan,Dolan2,Dolan3,Kubiznak,Cai,
Kubiznak2,Xu,Xu2,Zhm}.
The major innovation made in this stage is the inclusion of the 
negative cosmological constant (which is associated to the pressure), together with 
a conjugate variable known as the thermodynamic volume, as a new pair of 
thermodynamic variables. The sacrifice paid for this change is the reinterpretation  
of the mass parameter as the enthalpy. In this formalism,
very rich thermodynamic behaviors have been found for 
various black holes in a diverse number of gravity models. In particular,
phase transitions, either of the normal type (e.g. like in the Van der Waals 
liquid-gas system \cite{Kubiznak}) or of some exotic types (e.g. with multiple critical 
points, allowing for reentrant (zero-th order) and/or supercritical phase transitions,
{\em etc}. \cite{Xu, Xu2, Zhm}), were studied extensively. Such studies also triggered 
the motivation for including other model parameters as novel thermodynamic parameters 
\cite{Cai,Xu}. Meanwhile, the inclusion of the $(P,V)$ variables makes it possible 
to consider AdS black holes as heat engines \cite{Johnson,Xu3}. 
Most recently, Visser \cite{Visser} made a novel contribution to the study of 
EPST by including of the central charge $C$ and 
the chemical potential $\mu$ as a new pair of conjugate thermodynamic variables.
In this way, the first law of black hole thermodynamics is further extended, and the 
study of the corresponding central charge criticality has just emerged
\cite{CKM,Alfaia}. Please note that the idea of introducing a chemical potential 
which is conjugate to the central charge is explored earlier in \cite{Kastor2}. 
Similar ideas are also introduced in \cite{Zhang,Karch,Maity,Wei}, 
however the conjugate variable of the chemical potential 
is not the central charge but the square of number of colors in the dual theory,
which is related but not always equal to the central charge \cite{Karch,Rafiee:2021hyj}.

Despite the aforementioned significant progresses, there are still some aspects in 
black hole thermodynamics which remain poorly understood, and some of the issues 
are important enough to raise concerns about the overall completeness or consistency 
of the existing formalisms. 

In this paper, we will provide a new formalism for 
black hole thermodynamics which avoids most of the issues that appear in previous 
formalisms. In order to clarify our motivation and lay down the problem, 
we will briefly outline the major structure of the EPST and introduce Visser's 
improvement above it in Section 2. In this process, the existing issues and their 
seriousness will be made clear. Then, in Section 3, we develop the new formalism 
which we call the {\em restricted phase space thermodynamics} (RPST) for AdS black 
holes and examine its structure and merits in great detail in the example case of 
Reissner-Nordstr\"om anti-de Sitter (RN-AdS) black hole 
in 4-dimensional Einstein-Maxwell theory. Finally, in Section 4, we summarize the 
results and discuss about some possible future developments.

\section{Outline of the EPST and related issues}

We begin by briefly outline the structure of the EPST and clarify its major issues. 
For this purpose, it suffices to consider a rotating charged AdS black hole solution 
in $d$-dimensional Einstein-Maxwell theory as an illustrative example. 
The action of this theory is given by
\begin{align}\label{eq1.5_1}
\mathcal{A} &= \frac{1}{2\kappa_d}\int_{\mathcal{M}}\rd^d x\sqrt{-g}(R-2\Lambda)-\frac{1}{4\mu_0}\int_{\mathcal{M}}\rd^d x\sqrt{-g}(F_{\mu\nu}F^{\mu\nu}),
\end{align}
where $\kappa_d$ is the $d$-dimensional Einstein constant, $\mathcal{M}$ 
represents the spacetime manifold, and $F_{\mu\nu}=\nabla_{\mu} A_{\nu}
-\nabla_{\nu} A_{\mu}$ is the Maxwell field strength tensor.

The EPST works only in the case with a negative cosmological constant $\Lambda$. 
It extends the TBHT in that the {\em pressure} $P=-\frac{\Lambda}{8\pi G}$ 
associated with  $\Lambda$ is introduced as a new macro state parameter, 
and a conjugate variable, $V = r_H^{d-1} \mathcal{A}_{d-1}$, where $r_H$ represents 
the radius of the event horizon and $\mathcal{A}_{d-1}$ is the 
area of the unit $(d-1)$-sphere, commonly referred to as 
the thermodynamic volume, is also included, so that the first law in the EPST takes the 
form
\begin{align}
\rd M = T\rd S+\Omega\rd J+ \Phi\rd Q +V\rd P,
\label{epst1st}
\end{align}
where $M$ is the black hole mass, $T,S$ respectively represent the temperature and 
entropy, $\Omega, J$ are the angular velocity and angular momentum, 
and $\Phi, Q$ are the Coulomb potential and electric charge. 
Besides the first law, another important relation in the EPST 
is the Smarr relation,
\begin{align}\label{eq1.5_2}
M&= \frac{d-2}{d-3}\left(TS+ \Omega J\right) +\Phi Q-\frac{2}{d-3}PV.
\end{align}

The first issue of the EPST comes about by allowing $P$ to vary. Since the 
cosmological constant is among the model-defining parameters, {\em varying $P$ 
implies changing the underlying gravity model}. Therefore, the corresponding 
ensemble does not describe the collection of black holes from the same gravity 
model which are in the same macro state (i.e. {\em ensemble of macro systems}), 
but rather describes the collection of gravity models which admit 
the same or similar black hole solutions (i.e. {\em ensemble of gravity theories}). 
The standard excuse for this ``ensemble of theories'' picture is that there is 
some {\em underlying fundamental theory} of which the model parameters arise as 
moduli parameters. Given that the existence of the assumed fundamental theory is 
still unjustified, it feels more comfortable and safe to consider only the 
ensemble of black holes from the same gravity model instead of the ensemble
of gravity models admitting the same black hole solution. 

Several other issues will arise by simply looking at eq.\eqref{epst1st}. 
For instance, the appearance of the term $V\rd P$ 
implies that the mass $M$ is to be understood as enthalpy 
rather than internal energy. \blue{However, this understanding seems to be 
in contradiction with the definition of the black hole mass as the total 
energy of the black hole spacetime, or as the conserved charge associated with the 
timelike Killing vector field.}
Meanwhile, the meaning of the thermodynamic volume $V$ remains obscure, except that 
$V$ is required in order to write down the first law in the EPST. 

The last and the most important issue can be revealed by making comparison between eqs.\eqref{epst1st}
and \eqref{eq1.5_2}. From eq.\eqref{epst1st} one obtains
\[
T=\pfrac{M}{S}_{J,Q,P},\quad \Omega=\pfrac{M}{J}_{S,Q,P},\quad
\Phi=\pfrac{M}{Q}_{S,J,P},\quad V=\pfrac{M}{P}_{S,J,Q}.
\]
Inserting these relations into eq.\eqref{eq1.5_2} yields
\[
M=\frac{d-2}{d-3}\left[S \pfrac{M}{S}_{J,Q,P}+ J \pfrac{M}{J}_{S,Q,P}\right]
+Q \pfrac{M}{Q}_{S,J,P} -\frac{2}{d-3}V \pfrac{M}{P}_{S,J,Q}.
\]
This equation implies that $M$ is a homogeneous function of $S$ and $J$ 
with order $\frac{d-2}{d-3}$, of $Q$ with order $1$, and of $-P$ with order 
$\frac{2}{d-3}$. However, in standard thermodynamics, the thermodynamic potentials 
must be homogeneous functions of the related additive extensive variables of 
the {\em first order}, which is best exemplified by the Euler relation for the 
internal energy $E$ of an ordinary gaseous system consisting of a single constituent, 
\[
E=TS-PV+\mu N.
\]
It is very important that {\em the absolute values of the coefficients appearing on the RHS 
of the Euler relation must be uniformly $+1$}. This characteristic property 
\blue{has been realized neither in the traditional formalism nor in the EPST formalism}
\footnote{\blue{However it has indeed been realized in a particular formalism
developed specifically for AdS black holes in \cite{Tian1,Tian2}, see the notes added 
in the end of this paper.}}.

Recently, Visser \cite{Visser} made an important step towards the resolution of the 
last issue. The key idea is to take the central charge $C=L^{d-2}/G$ of the 
holographic dual theory as a new extensive variable, and consider the  
mass $M$ as a function of the variables $A, J, Q, \Lambda, G$, where $A$  
represents the area of the event horizon. It is observed that 
\begin{align}\label{eq1.5_2.5}
\rd M&=\frac{\kappa}{2\pi}\rd\bigg(\frac{A_{d-2}}{4G}\bigg)
+\Omega \rd J+\frac{\Phi}{L} \rd(QL)-\frac{M}{d-2}\frac{\rd L^{d-2}}{L^{d-2}} 
\nonumber\\
&\quad +\left(M-\frac{\kappa A_{d-2}}{8\pi G}-\frac{\Phi}{L} QL-\Omega J\right)
\frac{\rd(L^{d-2}/G)}{L^{d-2}/G},
\end{align}
where $L^2=-\frac{(d-1)(d-2)}{2\Lambda}$ is the AdS radius,
$G$ is Newton's constant in $d$-dimensions and $\kappa$ represents 
the surface gravity. By identifying\footnote{There is a slight difference between 
the conventions of \cite{Visser} and the present work: $\tilde \Phi$, $\tilde Q$
in eq.\eqref{identi} (which is cited from \cite{Visser}) should be changed into 
$\hat \Phi =\frac{\Phi \sqrt{G}}{L}$, $\hat Q=\frac{Q L}{\sqrt{G}}$ under 
our convention. In contrast, the convention used in \cite{CKM} agrees with ours.}
\begin{align}
E=M,\quad T=\frac{\kappa}{2\pi},\quad S=\frac{A_{d-2}}{4G},\quad
\tilde \Phi=\frac{\Phi}{L},\quad \tilde Q=QL,\quad V=L^{d-2}
\label{identi}
\end{align}
and introducing the equation of states \blue{(EOS)}
\begin{align}
E=(d-2)PV \label{eqm}
\end{align}
for the dual CFT\footnote{\blue{For $d$-dimensional AdS black hole spacetimes, the dual
CFT lives in a $(d-1)$-dimensional boundary spacetime, for which the EOS takes the form
\eqref{eqm}. One can imagine that eq.\eqref{eqm} is analogous to the EOS $E=3PV$ for a
photon gas in 4-dimensional spacetime, where 4 corresponds to $d-1$.}}, 
eq.\eqref{eq1.5_2.5} can be rearranged in the form
\begin{align}
\rd E= T\rd S+\Omega\rd J+\tilde \Phi\rd\tilde Q- P\rd V+\mu\rd C,	
\label{V1stL}
\end{align}
where $\mu$ is {\em defined} using the following identity,
\begin{align}
E= TS+ \tilde \Phi \tilde Q + \Omega J + \mu C.
\label{hEuler}
\end{align}
The rescaled electric charge $\tilde Q$ and Coulomb potential $\tilde \Phi$ can be 
understood as the corresponding quantities of the dual CFT. 
Since we encounter two charges $\tilde Q$ and $C$ in this context, we shall
call them the $e$-charge and the $C$-charge for short.

It is remarkable that eq.\eqref{hEuler} looks very similar to the Euler relation 
in standard thermodynamics, except that the $(P,V)$ variables are absent on the RHS. 
The meaning of $P, V$ has changed in this holographic description: $V$ is now 
a measure of the spatial size of the dual CFT (which has one less spatial dimension
comparing to the thermodynamic volume in the EPST), and $P$ is 
introduced in eq.\eqref{eqm} in order that the corresponding equation of states 
indeed describes a CFT. Since the $C$-charge is a measure for the number of 
microscopic degrees of freedom in CFT, its conjugate $\mu$
naturally acquires an explanation as the chemical potential. 
Notice that, since $V$ and $C$ are regarded as independent extensive 
variables, varying $C$ implies varying Newton's constant $G$. 
Another point to be noticed is the appearance of the term $-P\rd V$ in eq.\eqref{V1stL}. 
This term reflects the fact that $E$ restored its role as internal energy, and thus 
the first issue mentioned above is avoided. However, since $V$ is now defined in
terms of the AdS radius $L$, varying $V$ implies varying $\Lambda$.
Therefore, Visser's formalism still suffers from the ``ensemble of theories'' 
issue.

\section{RPST for RN-AdS black hole}

Recall that the AdS/CFT duality is a one to one 
correspondence between a given gravity model and the dual CFT. When considering 
a concrete AdS/CFT duality, it is important to fix a gravity model in the first place,  
rather than considering a collection of different gravity models 
altogether. This requirement forbids the appearance of the term $V\rd P$
in eq.\eqref{epst1st} and the term $-P\rd V$ in eq.\eqref{V1stL}. 
The formalism for black hole thermodynamics to be described below is free from 
the $(P,V)$ extension, and hence we call it the RPST. This formalism is free 
of all the issues mentioned in the last section. 

The simplest and most straightforward realization of the RPST is 
given by fixing $L$ in eqs.\eqref{V1stL} and \eqref{hEuler}, which results in
\begin{align}
\rd M&= T\rd S+\Omega\rd J+ \tilde\Phi\rd \tilde Q+\mu\rd C, 
\label{R1st}\\
M&= TS+ \tilde\Phi \tilde Q + \Omega J + \mu C.
\label{REuler}
\end{align}
These equations bear the standard form for the first law and Euler relation
in traditional thermodynamics. Since the variables $P,V$ no longer appear, 
there can be no volume work involved in black hole 
thermodynamic processes, hence the black hole cannot act as heat engine 
in this formalism. Even though, being a thermodynamic system with multiple 
thermodynamic degrees of freedom, the black hole can still have interesting 
thermodynamic behaviors in the RPST, as will be exemplified below by the 
RN-AdS black hole in 4 dimensions.

\blue{Since the RPST formalism still allows $G$ to vary, it seems necessary to 
pay some words in explaining the different roles of variable $G$ and 
variable $\Lambda$. Basically, $\Lambda$ is a part of the Lagrangian density, varying 
$\Lambda$ leads to a different set of gravitational field equations. On the contrary, 
$G$ can be arranged to become an overall factor in front of the total action,
the only sacrifice to be paid is a rescaling of the integration constant $Q$ in the 
case of Einstein-Maxwell theory. This not only explains the necessity of introducing a 
rescaling for the $e$-charge (and consequently also for the electric potential) 
in the subsequent analysis, but also ensures that varying $G$ 
does not lead to different field equations. That is why we think of our formalism to be 
free of the ``ensemble of theories'' issue. Another difference between variable 
$G$ and $\Lambda$ lies in that varying $\Lambda$ changes the geometry, while varying $G$ 
does not. We thank Visser for pointing out this last difference in private communications.
}

Before presenting the detailed study of the RN-AdS example, it is interesting to mention 
that an alternative restricted version of Visser's 
formalism \cite{Visser} with fixed $C$ but varying $V$ is studied in detail in 
\cite{Rafiee:2021hyj}, wherein the varying cosmological
constant is attributed to varying the curvature radius of the spacetime on which the dual CFT
resides. Here, our concentration is on the black hole side, with emphasis on the 
correct Euler relation and the 
homogeneity behaviors of the black hole mass and other state variables.

\subsection{Realization of the RPST for RN-AdS black hole}

To be more concrete, let us write down the metric of 4-dimensional RN-AdS black hole
together with its accompanied electromagnetic potential as follows,
\begin{align*}
&\rd s^2 = - f(r) \rd t^2 + f(r)^{-1}\rd r^2
+r^2(\rd\theta^2 +\sin^2\theta \rd\phi^2),\\
& A^\mu=\left(\Phi(r),0,0,0\right),
\end{align*}
where 
\[
f(r)=1- \frac{2GM}{r}+\frac{GQ^2}{r^2}+\frac{r^2}{\ell^2},
\qquad
\Phi(r)=\frac{Q}{r}.
\]
Here we work in units $\mu_0=4\pi, \epsilon_0=1/4\pi, c=\hbar=k_{\rm B}=1$, 
however keep $G$ untouched. $\ell$ is related to the cosmological constant via
$\displaystyle \ell^2=-\frac{3}{\Lambda}$. When $M>Q/\sqrt{G}$, $f(r)$ has
two distinct real zeros $r=r_{\pm}$, and the bigger one $r_+$ corresponds 
to the event horizon. The black hole mass $M$ can be described as a function of 
$r_+, G, Q$ by solving the equation $f(r_+)=0$, i.e.
\begin{align}
M=\frac{r_+}{2G}\left(1+ \frac{GQ^2}{r_+^2}+\frac{r_+^2}{\ell^2}\right).	
\label{Morig}
\end{align}

In the RPST for RN-AdS black hole, the macro states are characterized by the 
following three pairs of conjugate variables: $(S,T)$, $(\hat Q,\hat\Phi)$ and $(C,\mu)$.
The values for each pair of variables are given as follows.
First, the entropy $S$ and its conjugate, the temperature $T$, are given as
\begin{align}
S&=\frac{A}{4G}=\frac{\pi r_+^2}{G},\qquad
T=\frac{f'(r_+)}{4\pi}
=\frac{1}{2\pi r_+}\bigg(\frac{GM}{r_+}
-\frac{G Q^2}{r_+^2}+\frac{r_+^2}{\ell^2}\bigg).
\label{T-preEOS}
\end{align}
Next, the (rescaled) $e$-charge $\hat Q$ and it conjugate, the rescaled 
Coulomb potential $\hat \Phi$ on the event horizon, are given by
\begin{align}
\hat Q= \frac{Q\ell}{\sqrt{G}},\qquad	
\hat \Phi=\frac{\Phi(r_+)\sqrt{G}}{\ell}=\frac{Q\sqrt{G}}{r_+\ell}.
\label{QPhi}
\end{align}
Finally, the $C$-charge $C$ and its conjugate, the chemical potential $\mu$, are given by
\begin{align}
C=\frac{\ell^2}{G},	\qquad 
\mu= \frac{M-TS-\hat\Phi \hat Q}{C}.
\label{Cmu}
\end{align}
Inserting eq.\eqref{Morig} into eq.\eqref{T-preEOS} and then into \eqref{Cmu}, it becomes 
evident that the state variables $S,\hat Q, C$ are also
functions of $(r_+, G, Q)$. By straightforward calculations,
it can be verified that
\begin{align}
\rd M =T\rd S+\hat \Phi\rd\hat Q+\mu\rd C.	
\label{RN1st}
\end{align}
Moreover, the definition for $\mu$ given in eq.\eqref{Cmu} indicates that
\begin{align}
M=TS+\hat\Phi\hat Q	+\mu C.
\label{RNEuler}
\end{align}
Eqs.\eqref{RN1st}, \eqref{RNEuler} are nothing but a special case of 
eqs.\eqref{R1st}, \eqref{REuler} with $\Omega=J=0$. 
Using eqs.\eqref{RN1st} and \eqref{RNEuler}, we can easily get the 
Gibbs-Duhem relation
\begin{align}
\rd\mu=-\hat{\mathcal Q} \rd\hat\Phi -\mathcal{S} \rd T,
\label{GD}
\end{align}
where $\hat{\mathcal Q}=\hat Q/C$ and $\mathcal{S}=S/C$ are respectively the 
$e$-charge per unit $C$-charge and the entropy per unit $C$-charge. 
This equation is absent in both the TBHT and the EPST formalisms for AdS black holes.

\subsection{Equations of states and extensivity of $S$ and $\hat Q$}

In order to describe RN-AdS black hole in the RPST,
we need to rewrite the mass $M$ as a function of the variables $S,\hat Q$ 
and $C$. To do so, we first solve $G$ and $r_+$ in terms of $S,C$ using
the first equalities in eqs.\eqref{T-preEOS} and \eqref{Cmu},
\begin{align}
G=\frac{\ell^2}{C},\qquad 
r_+=\ell\left(\frac{S}{\pi C}\right)^{1/2}.
\label{Gr}
\end{align}
Then, substituting eq.\eqref{Gr} into \eqref{Morig}, we get the desired result,
\begin{align}
M(S,\hat Q,C)=\frac{{S}^{2}+\pi S C+ {\pi}^{2}{\hat Q}^{2}}{2{\pi}^{3/2}{\ell}(SC)^{1/2}}.
\label{Mext}
\end{align}
The first law \eqref{RN1st} allows us to write down the equations of states 
directly as partial derivatives of $M$,
\begin{align}
T&=\pfrac{M}{S}_{\hat Q,C}
=\frac {3{S}^{2}+\pi S C-\pi^2\hat Q^2}
{4{\pi}^{3/2}\ell S(SC)^{1/2}},
\label{Text}
\\
\hat \Phi&=\pfrac{M}{\hat Q}_{S,C}
=\left(\frac{\pi}{SC}\right)^{1/2}\frac{\hat Q}{\ell},
\label{Phiext}
\\
\mu&=\pfrac{M}{C}_{S,\hat Q}
=-\frac{S^2-\pi SC+\pi^2 \hat Q^2}{4{\pi}^{3/2}\ell {C}(SC)^{1/2}}.
\label{muext}
\end{align}
If $S,\hat Q,C$ are rescaled as $S\to \lambda S, \hat Q\to\lambda \hat Q, C\to\lambda C$,
then eq.\eqref{Mext} tells that $M$ will also rescale as $M\to\lambda M$, while 
eqs.\eqref{Text}-\eqref{muext} indicate that $T, \hat\Phi, \mu$ will not get rescaled. 
Thus the first order homogeneity of $M$ and the zeroth order homogeneity of 
$T, \hat\Phi, \mu$ are clear as crystal. Please note that, the zeroth order 
homogeneous functions are, by definition, intensive.

It remains to check that the variables $S$ and $\hat Q$ are extensive. 
This is a necessary step because the area dependence of the black hole entropy 
is often missinterpreted as nonextensivity. 

First we need to make clear what is meant by the term extensivity here. 
According to the first law \eqref{RN1st}, the $C$-charge plays the role like 
the particle number, therefore it is automatically extensive. Other extensive variabes 
are defined such that they are proportional to $C$, with the coefficients 
of proportionality depending only on the intensive variables. 

Now let us prove that $S$ is extensive in the above sense. To do so, let us first 
solve $\hat Q$ from eq.\eqref{Phiext},
\begin{align}
\hat Q= \ell\hat\Phi\left(\frac{SC}{\pi}\right)^{1/2}.
\label{QvsPhi}
\end{align}
Substituting eq.\eqref{QvsPhi} into eq.\eqref{Text}, we get
\begin{align}
T&=\frac {3{S}+\pi C\left( 1-\ell^2 \hat \Phi^{2}\right) }
{4{\pi}^{3/2}\ell (SC)^{1/2}}.
\label{TPhiSC}
\end{align}
In order that $T$ takes a non-negative value, it is necessary to constrain the 
value of $\hat\Phi$, so that for any $S\geq 0$, 
we have $T\geq 0$. This leads to the following bound
\begin{align}
\hat \Phi \leq \frac{1}{\ell}.
\label{bPhi}
\end{align}
Assuming that the above bound is unsaturated, we can solve $S$ out of eq.\eqref{TPhiSC}.
The result reads
\begin{align}
S&=\mathcal{S}C,\qquad
\mathcal{S}
= \frac{\pi^3\ell^2}{9}\left[8\tau^2-3\pm 4\tau\left(4\tau^2-3\right)^{1/2}\right]T_0^2,
\label{SC}
\end{align}
where 
\begin{align}
\tau=\frac{T}{T_0},\qquad
T_0\equiv \frac{\left(1-\ell^2\hat\Phi^2\right)^{1/2}}{\pi\ell}.
\label{taut0}
\end{align}
We see that $S$ is indeed proportional to $C$, with the coefficient $\mathcal{S}$ 
(which already appeared in eq\eqref{GD}) depending only on the intensive variables 
$T$ and $\hat\Phi$. This proves the extensivity of $S$. By the way, 
the double-valuedness of $S$ indicated by eq.\eqref{SC} will be explained 
in the next subsection.

The proof for extensivity of $\hat Q$ is accomplished by inserting eq.\eqref{SC} 
into \eqref{QvsPhi}. The result reads 
\begin{align*}
\hat Q=\mathcal{\hat Q}C,\qquad 
\mathcal{\hat Q}= \ell \hat\Phi \left(\frac{\mathcal{S}}{\pi}\right)^{1/2},
\end{align*}
where $\mathcal{\hat Q}$ depends solely on the intensive variables $T,\hat\Phi$.

\subsection{Thermodynamic processes and phase transitions}

Eqs.\eqref{Text}-\eqref{muext} provide three constraint conditions over 
six variables. Therefore, a black hole macro state is characterized by only three of the 
thermodynamic variables $T,\hat \Phi, \mu, S, \hat Q$ and $C$. 
If any of the three chosen variables varies, a macroscopic process takes place. 
Notice that in eqs.\eqref{Mext}, \eqref{Text} and \eqref{muext},
$\hat Q$ always appear in squared form. The only exception appears in 
eq.\eqref{Phiext}, which indicates that $\hat \Phi$ and $\hat Q$ always take the same 
signature. Therefore, it suffices to consider only the cases
with $\hat Q\geq0$.

Since every intensive variables depends on three extensive ones
in the equations of states, the allowed thermodynamic 
processes can be extremely complicated. For simplicity, we only consider processes
along certain simple curves by fixing two of the independent variables.

First let us consider the $T-S$ curves at fixed $\hat Q, C$. 
The extremal black hole state with $T=0$ is approached if
\[
\hat Q=\hat Q_{\rm max}=\frac{\left[S(\pi C+3S)\right]^{1/2}}{\pi}.
\]
This gives an upper bound for the value of $\hat Q$ at each choices of $S,C$.
Assume that the bound $\hat Q\leq \hat Q_{\rm max}$ is unsaturated. 
We wish to find the inflection point on the $T-S$ curve with fixed $\hat Q, C$ 
which is characterized by the equations
\begin{align*}
\pfrac{T}{S}_{\hat Q, C}=0,\qquad \pfn{T}{S}{2}_{\hat Q, C}=0.
\end{align*}
It turns out that the inflection point exists only when $\hat Q$ takes the 
critical value
\begin{align*}
\hat Q=\hat Q_c \equiv \frac{{C}}{6},
\end{align*}
wherein $C$ can take any positive value. The critical values for $S$ and $T$ are
given by
\begin{align*}
S_c= \frac{\pi C}{6},\qquad T_c=\frac{\sqrt{6}}{3\pi \ell}.
\end{align*}
Introducing the relative parameters
\begin{align*}
t=\frac{T}{T_c},\quad s=\frac{S}{S_c},\quad q=\frac{\hat Q}{\hat Q_c},	
\end{align*}
eq.\eqref{Text} can be rewritten as
\begin{align}
t=\frac{3s^2+6s-q^2}{8s^{3/2}}.	
\label{relT}
\end{align}
It is interesting to notice that the $T-S$ EOS rewritten in the form \eqref{relT} does not 
depend on the value of $C$. This is typical in the study of thermodynamics of ordinary matter,
known as the law of corresponding states.

We can also introduce the Helmholtz free energy by a Legendre transform of eq.\eqref{Mext},
\begin{align*}
F(T,\hat Q,C)=M(S,\hat Q,C)-TS,
\end{align*}
which, in terms of the relative parameter $f=F/F_c$ with $F_c=\frac{\sqrt{6}C}{18\ell}$, 
is given as
\begin{align}
f(t,q)=\frac {{q}^{2}+{s}^{2}+6s-4\,t{s}^{3/2}}{4{s}^{1/2}}.
\label{relF}
\end{align}
We see once again that the free energy equation can be written in a form which is independent of $C$.
This shows the power of scaling properties in thermodynamics. 
Please note that, in eq.\eqref{relF}, the variable $s$ appearing in eq.\eqref{relF} must be 
understood as being implicitly determined by $t$ and $q$ via eq.\eqref{relT}.

\begin{figure}[ht]
\begin{center}
\includegraphics[width=.48\textwidth]{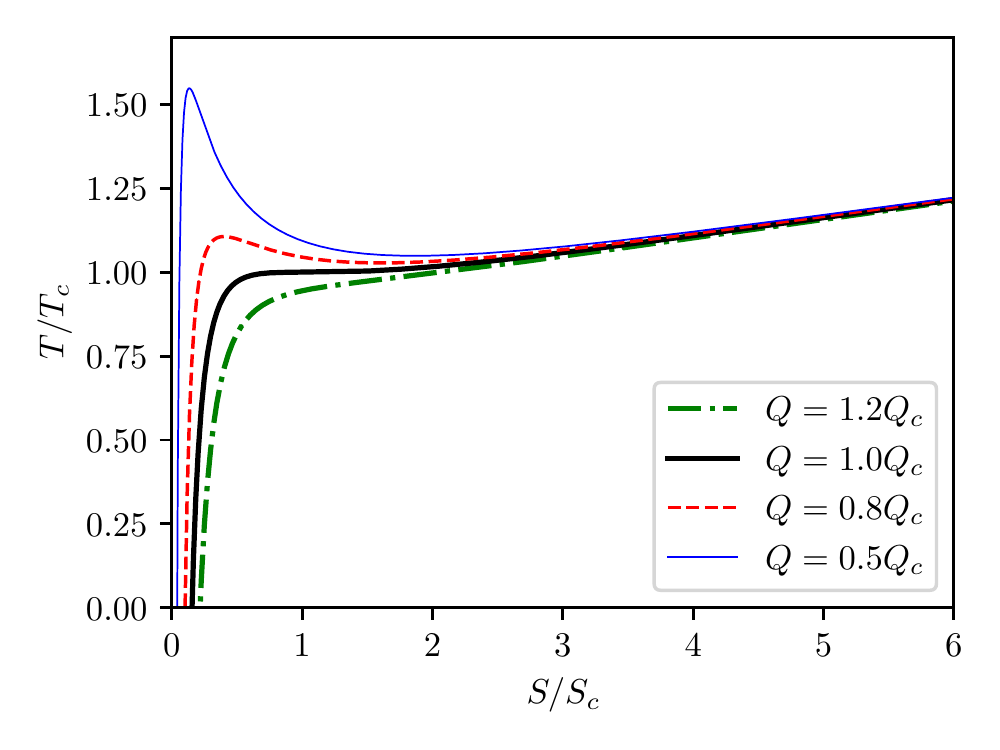}\hspace{2pt}
\includegraphics[width=.48\textwidth]{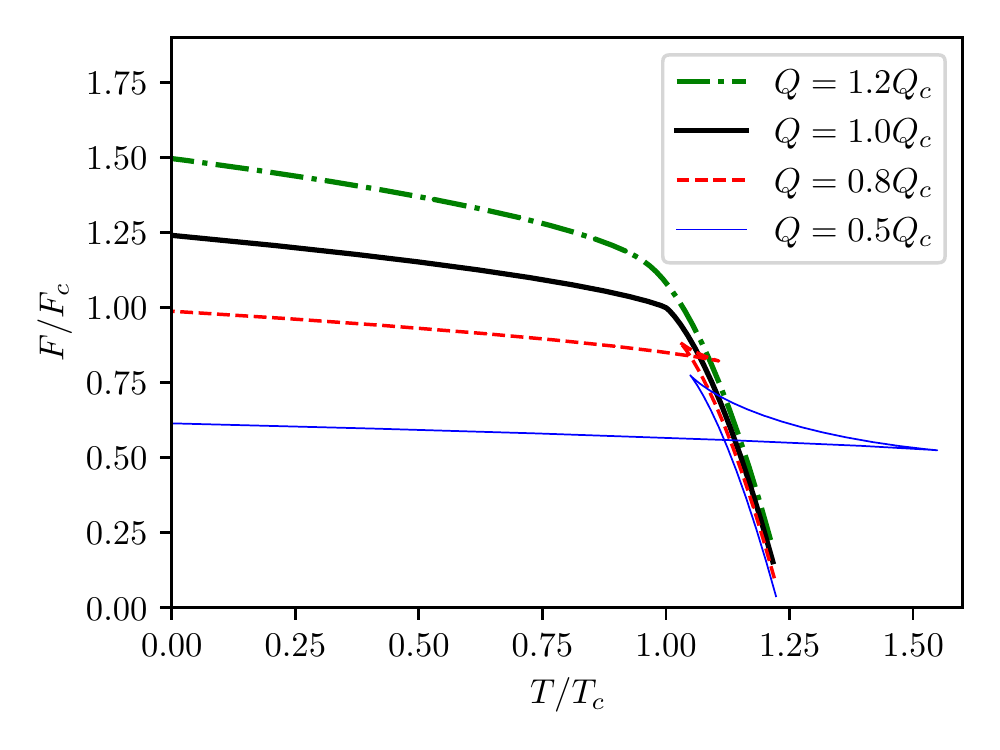}
\caption{$T-S$ and $F-T$ curves in the iso-$e$-charge processes}\label{fig1}
\end{center}
\end{figure}

The $T-S$ and $F-T$ curves corresponding to the iso-$e$-charge processes can be 
depicted using eqs.\eqref{relT} and \eqref{relF}, as shown in Fig.\ref{fig1}. 
It can be seen that, 
at any {\em subcritical value} of $\hat Q$, the $T-S$ curve becomes 
non-monotonic, correspondingly, the $F-T$ curve contains a swallow tail with the root  
located at some {\em supercritical temperature} $T>T_c$. This behavior indicates that 
there is a Van der Waals like first order phase equilibrium appearing in the 
iso-$e$-charge processes provided $0<\hat Q<\hat Q_c$. 
Since we did not associate the black hole with the concept of a volume,
the above process is {\em not} accompanied by any volume work. If
$\hat Q=\hat Q_c$, the phase transition can still occur, but with the order 
shifted into the second. 

That the first order phase transition occurs 
at supercritical temperatures is something to be noticed. Such phenomena also 
appear in the EPST formalism for black hole thermodynamics, but never appear 
in the thermodynamics of ordinary matter.  
Another point to be noticed in Fig.\ref{fig1} is that, for any $\hat Q>0$, 
the iso-$e$-charge $T-S$ curve intersects with the $S$ axis at some nonvanishing $S=S_0$. 
This means that, when the 
black hole carries out a (not necessarily spontaneous) iso-$e$charge process and get its 
temperature decreased, the process must stop at $S=S_0$, which corresponds to 
a black hole remnant, from where the entropy cannot be further decreased. 
If $\hat Q=0$, the iso-$e$-charge $T-S$ curve will not intersect 
with the $S$ axis. Instead, $T$ acquires a single extremum which is a minimum 
at some nonvanishing value of $S$, 
which indicates that the phase transition becomes one of a different type, but we will not
expand on this special case because similar phenomenon will also happen  
in the isovoltage processes regardless of the $\hat Q$ values,
as will be discussed below.

Let us consider the isovoltage processes by depicting the 
$T-S$ curve at fixed $\hat \Phi, C$, assuming that the bound \eqref{bPhi} 
is unsaturated. It can be seen that, at fixed $\hat \Phi$ 
and $C$, eq.\eqref{TPhiSC} regarded as a function $T=T(S)|_{\hat \Phi,C}$ does 
not have any inflection point. Rather, 
there is a single extremum which corresponds to the minimum of $T$ located at
\[
S=S_{\rm min} = \frac{\pi C}{3}(1-\ell^2\hat\Phi^2),\qquad
T=T_{\rm min}=\frac{\sqrt{3}}{2}\frac{\left(1-\ell^2\hat\Phi^2\right)^{1/2}}{\pi\ell}
=\frac{\sqrt{3}}{2}T_0.
\]
Defining 
\[
\tilde t= \frac{T}{T_{\rm min}},\qquad \tilde s= \frac{S}{S_{\rm min}},
\]
eq.\eqref{TPhiSC} can be recast in the simple form 
\begin{align*}
\tilde t=\frac{\tilde s+1}{2\tilde s^{1/2}}.
\end{align*}

\begin{figure}[ht]
\begin{center}
\includegraphics[width=.5\textwidth]{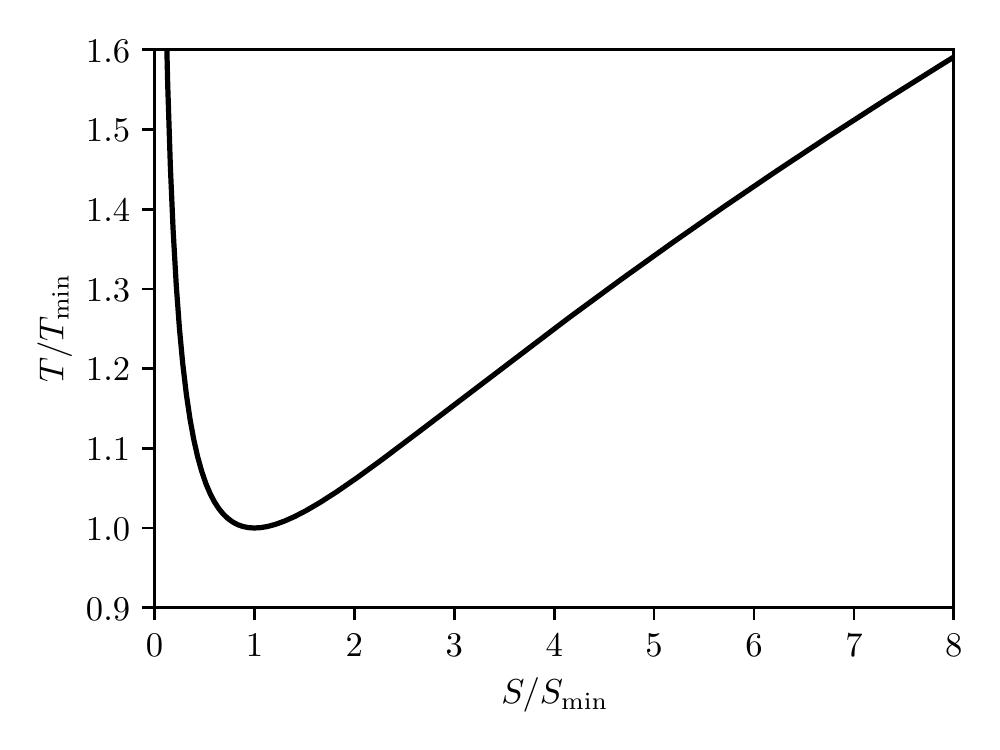}
\caption{$T-S$ curve in the isovoltage processes}\label{fig2}
\end{center}
\end{figure}
Fig.\ref{fig2} depicts the $T-S$ curve in the isovoltage process. 
Unlike the iso-$e$-charge $T-S$ curves, the isovoltage $T-S$ curve 
does not discriminate different $\hat\Phi$ values, because curves with different $\hat\Phi$
are scaled into a single one by use of the relative variables $\tilde s$, $\tilde t$. 
This is the law of corresponding states at an enhanced level.
The isovoltage process does not allow the temperature to be decreased 
below $T_{\rm min}$. At $T>T_{\rm min}$, there are two black hole states with the 
same $T,\hat\Phi,C$ but with different entropies (i.e. different horizon radii). 
This fact has already been reflected by eq.\eqref{SC}. 
According to the principle of maximum entropy, 
the state with bigger $S$ should be thermodynamically preferred. 
In other words, if the black hole in the state $(T,\hat\Phi,C)$ happens to have 
the smaller radius, it will jump to the one with the bigger radius under any small 
perturbation. This phenomenon should also be considered to be a phase transition, 
though there is not a specific transition temperature. Therefore, such 
phase transitions should be regarded as being non-equilibrium transitions.
The same conclusion can be achieved from a different perspective. 
Since $T$ decreases as $S$ increases for $S<S_{\rm min}$, the heat capacity
$C_{\hat\Phi,C}=T\pfrac{T}{S}_{\hat\Phi,C}<0$ for $S<S_{\rm min}$, therefore 
the states with $S<S_{\rm min}$ are unstable, whereas $C_{\hat\Phi,C}>0$ for $S>S_{\rm min}$,
thus the states with $S>S_{\rm min}$ are stable.

Remember that the above analysis works only when the bound \eqref{bPhi} is unsaturated.  
If, instead, the bound \eqref{bPhi} is saturated, 
the minimum on the $T-S$ curve will disappear, and the curve extends to 
the origin of the $(T,S)$ plane. In this case, $T$ increases monotonically
with $S$, the heat capacity $C_{\hat\Phi,C}$ is positive for any $S>0$, 
and the non-equilibrium phase transitions no longer appear.

It is tempting to consider also the thermodynamic processes along the
$\hat \Phi-\hat Q$ and $\mu-C$ curves. However, according to eq.\eqref{Phiext},
$\hat\Phi$ is proportional to $\hat Q$ for fixed $S,C$. Therefore, the adiabatic 
iso-$C$-charge process on the $(\hat \Phi,\hat Q)$ plane is trivial: 
there is neither inflection point nor extremum on the $\hat \Phi-\hat Q$ curve. 

On the other hand, from eq.\eqref{muext}
we find that there is a maximum for $\mu$ at fixed $S,\hat Q$ which is located at 
\[
C=C_{\rm max}\equiv \frac{3(S^2+\pi^2\hat Q^2)}{\pi S},\qquad
\mu=\mu_{\rm max}\equiv\frac{\sqrt{3}}{18\ell}\frac{S}{\left(S^2+\pi^2\hat Q^2\right)^{1/2}}.
\]
Introducing the dimensionless parameters 
\[
c=\frac{C}{C_{\rm max}},\qquad m=\frac{\mu}{\mu_{\rm max}},
\]
the $\mu-C$ equation of states can be rescaled into
\begin{align}
m=\frac{3c-1}{2c^{3/2}}.
\label{mcr}
\end{align}
Naturally, for the above rescaling of variables to make sense, the conditions 
$S\neq0$ and $S^2+\pi^2\hat Q^2\neq0$ must hold.

\begin{figure}[ht]
\begin{center}
\includegraphics[width=.5\textwidth]{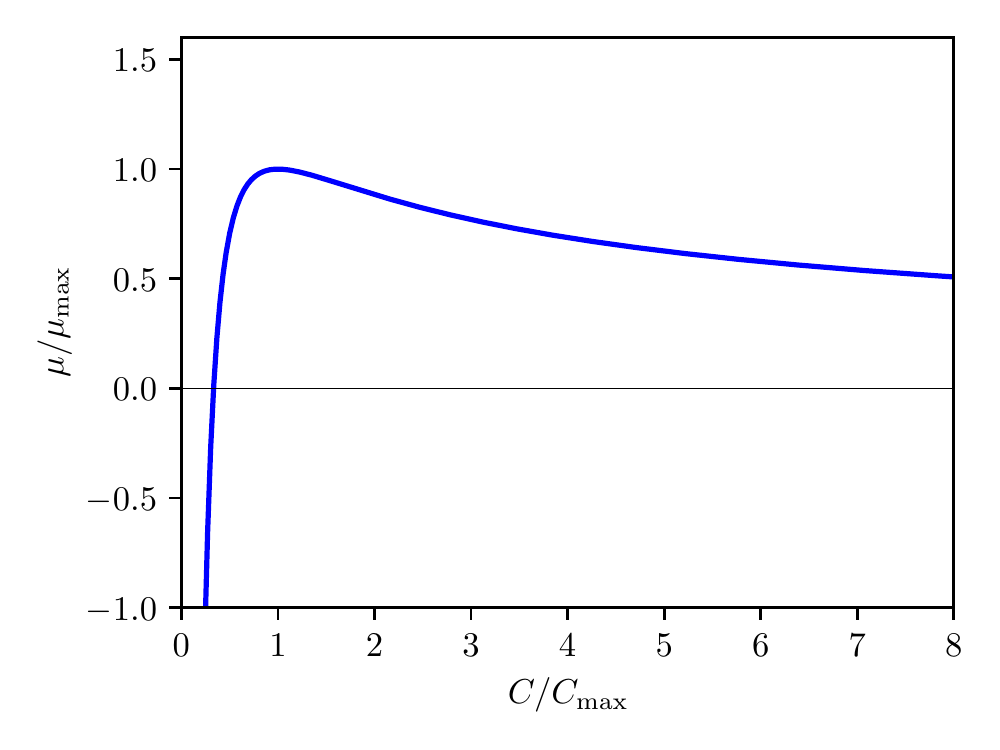}
\caption{$\mu-C$ curve in the adiabatic iso-$e$-charge/isovoltage processes}
\label{fig3}
\end{center}
\end{figure}

Fig.\ref{fig3} depicts the $\mu-C$ curve at fixed $S,\hat Q$. 
Just like the isovoltage $T-S$ curve, the $\mu-C$ curves for different choices of 
$S,\hat Q$ are all scaled into a single curve by use of the relative parameters $c, m$.
It is remarkable that at $C=\frac{1}{3}C_{\rm max}$, we have $\mu=0$, hence the Gibbs 
free energy $G=\mu C$ vanishes. 
For $C>\frac{1}{3}C_{\rm max}$, $\mu$ becomes positive but not too positive to exceed
$\mu_{\rm max}$. Therefore, 
\blue{at $C>\frac{1}{3}C_{\rm max}$, the microscopic degrees of freedom for the 
RN-AdS black hole are repulsive, while at $C<\frac{1}{3}C_{\rm max}$, they become attractive.}

One may also be interested in studying the $\mu-C$ processes at fixed $S$ and $\hat \Phi$.  
This can be done by replacing $\hat Q$ in eq.\eqref{muext} by use of eq.\eqref{Phiext}.
Omitting the details, we find that the $\mu-C$ relation with fixed $S,\hat\Phi$ 
can be reduced into the same equation \eqref{mcr}, but with the maximal values 
$C_{\rm max}$ and $\mu_{\rm max}$ represented in terms of $\hat\Phi$,
\[
C_{\rm max}= \frac{3S}{\pi(1-\ell^2\hat\Phi^2)},\qquad
\mu_{\rm max} 
= \frac{\sqrt{3}}{18\ell} \left(1-\ell^2\hat\Phi^2\right)^{3/2}.
\]
Therefore, Fig.\ref{fig3} can also be viewed as the adiabatic isovoltage
$\mu-C$ curve. As expected, the existence of the maximum of $\mu$ 
relies on the fact that the bound \eqref{bPhi} is unsaturated, or equivalently on the
fact that the black hole is non-extremal.

\begin{figure}[ht]
\begin{center}
\includegraphics[width=.5\textwidth]{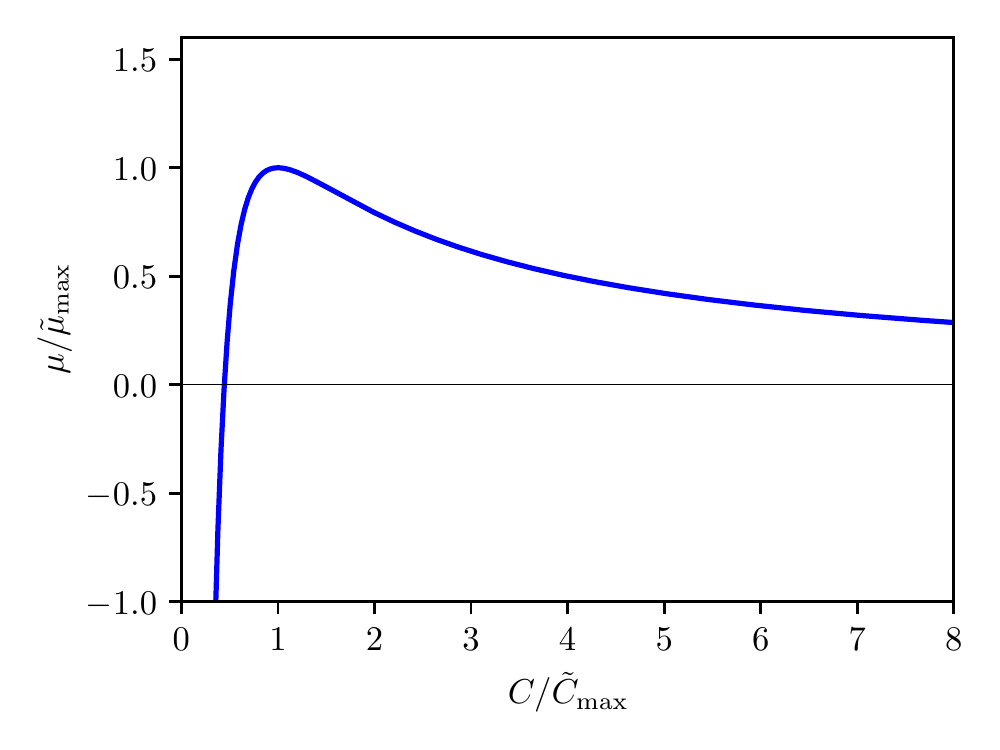}
\caption{$\mu-C$ curve in the adiabatic isothermal processes}\label{fig4}
\end{center}
\end{figure}

Besides isocharge $\mu-C$ processes, we may also consider the isothermal $\mu-C$ processes.
By use of eqs.\eqref{Text} and \eqref{muext}, we can find that 
\begin{align}
\mu=\frac{TS}{C}-\frac{S^2}{\pi^{3/2}\ell(SC)^{1/2}}.
\label{muCTS}
\end{align}
This equation is free of both the $e$-charge and the $C$-charge, but
involves $T$ and $S$. Therefore, we can make use of the above equation to study the 
adiabatic isothermal processes. It is easy to see that, at fixed $T,S$, eq.\eqref{muCTS}
has a single maximum which is located at
\[
\tilde C_{\rm max}=\frac{9S}{4\pi^3\ell^2 T^2},\qquad
\tilde \mu_{\rm max} =\frac{4\pi^3\ell^2 T^3}{27}.
\]
Using the dimensionless parameters 
\[
\tilde c=\frac{C}{\tilde C_{\rm max}},\qquad \tilde m=\frac{\mu}{\tilde \mu_{\rm max}},
\]
eq.\eqref{muCTS} becomes
\[
\tilde m=\frac{3\tilde c^{1/2}-2}{\tilde c^{3/2}}.
\]
The corresponding plot is given in Fig.\ref{fig4}, which looks very similar to 
Fig.\ref{fig3}, though it corresponds to different processes.

\blue{In either Fig.\ref{fig3} or Fig.\ref{fig4} one can see zeros of the chemical 
potential. In the absence of $e$-charge, such zeros correspond to the famous 
Hawking-Page (HP) transition, which corresponds to the transition from an AdS black 
hole state to a pure thermal AdS gas. Therefore, it is also interesting to analyze the 
HP transition in the present context.
}

An important concept related to HP transition is the HP
temperature $T_{\rm HP}$, i.e. the temperature at which HP transition occurs.
For the present case, $T_{\rm HP}$ can be worked out exactly. To solve $T_{\rm HP}$,
we \blue{simply set $\hat{Q}=0$ in} \eqref{muext}, yielding
\blue{
\begin{align}
\mu&= -\frac{S^{1/2}(S-\pi C)}{4{\pi}^{3/2}\ell {C}^{3/2}}.
\label{muSCPhi}
\end{align}}
Then, substituting eq.\eqref{SC} into eq.\eqref{muSCPhi}, we get
\begin{align}
\blue{\mu(T)}&= -\frac{\pi^3\ell^2}{27}(T_0)^{3}
\left[8\tau^2-3\pm 4\tau\left(4\tau^2-3\right)^{1/2}\right]^{1/2}
\left[2\tau^2-3\pm \tau \left(4\tau^2-3\right)^{1/2}\right],
\end{align}
where $\tau$ \blue{is} given in eq.\eqref{taut0} \blue{with $T_0$ replaced by
$\displaystyle T_0=\frac{1}{\pi\ell}$.} It is evident that  
the positive branch of \blue{$\mu(T)$} has a zero at $\tau=1$, which means that 
the sought-for HP temperature  $T_{\rm HP}$ is precisely $T_0$,
\[
T_{\rm HP}=T_0=\frac{1}{\pi\ell}.
\]
\blue{Moreover, according to eq.\eqref{muSCPhi}, the entropy at the HP transition point is 
$S=\pi C$, which, combined with the original 
Bekenstein-Hawking entropy expression given in eq. \eqref{T-preEOS} and the 
defining relation \eqref{Cmu} for $C$, reveals that the HP transition occurs precisely when
the horizon radius $r_+$ reaches the AdS radius $\ell$.}

\begin{figure}[ht]
\begin{center}
\includegraphics[width=.5\textwidth]{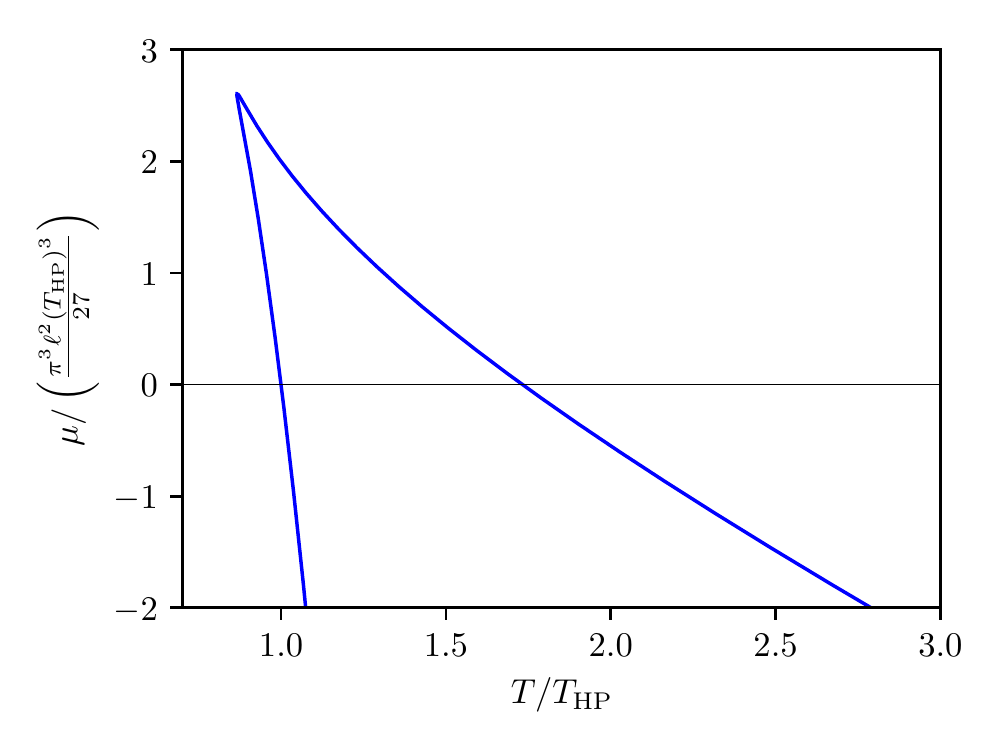}
\caption{The isovoltage $\mu-T$ curve}\label{fig5}
\end{center}
\end{figure}

We close this subsection by plotting the isovoltage $\mu-T$ curve in Fig.\ref{fig5}.
\blue{Please notice that the coexistence point of the two branches for $\mu(T)$ is located at 
a temperature below $T_{\rm HP}$. The HP transition occurs at the crossing point of the 
lower branch of $\mu(T)$ with the $\mu=0$ line.}

\section{Concluding Remarks}

\blue{A new formalism} for thermodynamics of asymptotically AdS black holes
with the correct Euler relation and Gibbs-Duhem equation is established. 
The resulting RPST is based on top of Visser's construction made in \cite{Visser}, 
but with an important restriction, i.e. the AdS radius must
be fixed. This new formalism is not only consistent with the usual logic
of AdS/CFT duality, but also removes the un-necessary volume work in the 
description of the first law. The mass parameter restores its role as internal energy,
and the thermodynamic processes do not lead the black hole solution of one 
underlying theory to that of another theory.

In spite of the simple algebra used while analyzing the example case of 4-dimensional RN-AdS 
black holes in Einstein-Maxwell theory, the significance of the results is 
highly nontrivial. It is shown that the entropy and the $e$-charge are both extensive in the sense 
that they are both proportional to the $C$-charge with the coefficients of proportionality
depending only on the intensive variables, and that the mass is a homogeneous function 
of the first order in all the extensive variables. 

It is also shown that there can be a first order Van der Waals like phase transition 
in the iso-$e$-charge processes on the $(T,S)$ plane if the $e$-charge $\hat Q$ takes some 
nonvanishing subcritical value $0<\hat Q<\hat Q_c=\frac{C}{6}$. If $\hat Q$ approaches the critical 
value, the phase transition becomes a second order one. However, when considering the 
isovoltage processes, the $T-S$ curve acquires a single minimum, 
which signifies a non-equilibrium phase transition. The constant $S,C$ 
process on the $(\hat \Phi,\hat Q)$ plane is trivial, and 
the constant $(S,\hat Q)$, constant $(S,\hat\Phi)$ and constant $(T,S)$ processes 
on the $\mu-C$ plane contains a single maximum and a Hawking-Page transition point. 
The study for the phase structure is far from being complete yet, but what we have done
already is sufficient to indicate that there are very 
rich thermodynamic structures in the RPST which worths further explorations.

As a final remark, let us stress that, since Visser's construction is fairly general, 
it is highly expected that the RPST should work for AdS black holes in generic 
gravity models. Exploring the applicability of the RPST and studying the 
corresponding thermodynamic behaviors in other gravity models are subjects of 
immediate interests. We hope to come back on these topics very soon.

\vspace{1em}

\noindent\blue{\em Notes added:} 

\blue{After the original version of this manuscript has been submitted,
we became aware that, an alternative formalism for AdS black hole
thermodynamics, which is intensively dependent on a holographic 
interpretation, has been introduced in Refs.\cite{Tian1,Tian2}, 
in which the Euler relation is also established,
but with a different set of extensive variables. The relationship between the formalism 
presented in \cite{Tian1,Tian2} and ours remains unclear. However, further studies 
reveals that our formalism works also for non-AdS black holes without 
a holographic interpretation \cite{Zhao3,Zhao4},
which justifies its wider applicability.
}

\section*{Acknowledgement}
This work is supported by the National Natural Science Foundation of 
China under the grant No. 11575088.

\providecommand{\href}[2]{#2}\begingroup
\footnotesize\itemsep=0pt
\providecommand{\eprint}[2][]{\href{http://arxiv.org/abs/#2}{arXiv:#2}}


\begin{thebibliography}{99}

\bibitem{Bekenstein} J. D. Bekenstein, ``Black holes and the second law,'' 
\href{https://link.springer.com/article/10.1007\%2FBF02757029}{\emph{Lett. Nuovo Cim.} 
4 (1972) 737–740}. 

\bibitem{Bekenstein2} J. D. Bekenstein, ``Black holes and entropy,'' 
\href{http://dx.doi.org/10.1103/PhysRevD.7.2333}{\emph{Phys. Rev. D} 7 
(1973) 2333–2346}. 

\bibitem{Bardeen} J. M. Bardeen, B. Carter, and S. W. Hawking, 
``The Four laws of black hole mechanics,''
\href{https://link.springer.com/article/10.1007\%2FBF01645742}{\emph{Commun. Math. Phys.} 
31 (1973) 161–170}.

\bibitem{Hawking} S. W. Hawking, ``Particle Creation by Black Holes,'' 
\href{https://link.springer.com/article/10.1007\%2FBF02345020}{\emph{Commun. Math. Phys.} 
43 (1975) 199–220}. [Erratum: Commun. Math. Phys. 46, 206 (1976)].

\bibitem{Hooft} G. ’t Hooft, ``Dimensional reduction in quantum gravity,''
[\eprint{gr-qc/9310026}].

\bibitem{Susskind} L. Susskind, ``The world as a hologram,'' 
\href{https://aip.scitation.org/doi/10.1063/1.531249}{\emph{J. Math. Phys.} 
36 (1995) 6377–6396}, [\eprint{hep-th/9409089}].

\bibitem{Maldacena} M. Maldacena, ``The large N limit of superconformal field 
theories and supergravity,'' \href{https://dx.doi.org/10.4310/ATMP.1998.v2.n2.a1}
{\emph{Adv. Theor. Math. Phys.} 2(1998) 231–252}.

\bibitem{Wald} R. M. Wald, ``Black hole entropy is noether charge,'' 
\href{https://journals.aps.org/prd/abstract/10.1103/PhysRevD.48.R3427}
{\emph{Phys. Rev.D} 48, R3427 (1993)} [\eprint{gr-qc/9307038}].

\bibitem{Hawking2} S. W. Hawking, D.N. Page, ``Thermodynamics of black holes in 
anti-de Sitter space,''
\href{https://link.springer.com/article/10.1007\%2FBF01208266}
{\emph{Commun. Math. Phys.} 87 (1983) 577}.

\bibitem{Kastor} D. Kastor, S. Ray, J. Traschen, ``Enthalpy and the mechanics of AdS 
black holes,'' \href{https://iopscience.iop.org/article/10.1088/0264-9381/26/19/195011}
{\emph{Class. Quant. Grav.} 26, 195011 (2009)},
[\eprint{0904.2765}].

\bibitem{Dolan} B. P. Dolan, ``The cosmological constant and the black hole 
equation of state,'' \href{https://iopscience.iop.org/article/10.1088/0264-9381/28/12/125020}
{\emph{Class. Quant. Grav.} 28, 125020 (2011)}, 
[\eprint{1008.5023}].

\bibitem{Dolan2} B. P. Dolan, ``Pressure and volume in the first law of black
hole thermodynamics,'' 
\href{https://iopscience.iop.org/article/10.1088/0264-9381/28/23/235017}
{\emph{Class. Quant. Grav.} 28, 235017 (2011)}, 
[\eprint{1106.6260}].

\bibitem{Dolan3} Dolan, ``Compressibility of rotating black holes,'' 
\href{https://doi.org/10.1103/PhysRevD.84.127503}
{\emph{Phys. Rev. D 84}: 127503 (2011)}, [\eprint{1109.0198}]

\bibitem{Kubiznak} D. Kubiznak, R.B. Mann, ``P-V criticality of charged AdS black holes,''
\href{https://link.springer.com/article/10.1007\%2FJHEP07\%282012\%29033}
{\emph{JHEP} 1207, 033 (2012)}, [\eprint{1205.0559}].

\bibitem{Cai} R.-G. Cai, L.-M. Cao, L. Li, and R.-Q. Yang, ``P-V criticality in 
the extended phase space of Gauss-Bonnet black holes in AdS space,''
\href{https://link.springer.com/article/10.1007\%2FJHEP09\%282013\%29005}
{\emph{JHEP} (2013) 005},[\eprint{1306.6233}]. 

\bibitem{Kubiznak2} D. Kubiznak, R. B. Mann, M. Teo, 
``Black hole chemistry: thermodynamics with Lambda,''
\href{https://iopscience.iop.org/article/10.1088/1361-6382/aa5c69}
{\emph{Class. Quantum Grav.} 34 063001 (2017)}, [\eprint{1608.06147}].

\bibitem{Xu} W. Xu, H. Xu, and L. Zhao, ``Gauss-bonnet coupling constant as a 
free thermodynamical variable and the associated criticality,'' 
\href{https://link.springer.com/article/10.1140\%2Fepjc\%2Fs10052-014-2970-8}
{\emph{The European Physical Journal C}, 74, No.7 (2014) 2970}, 
[\eprint{1311.3053}].

\bibitem{Xu2} W. Xu, L. Zhao, ``Critical phenomena of static charged AdS 
black holes in conformal gravity,''
\href{https://linkinghub.elsevier.com/retrieve/pii/S0370269314005218}
{\emph{Phys. Lett. B} 736 (2014) 214-220}, [\eprint{1405.7665}].


\bibitem{Zhm} M. Zhang, D. -C. Zou, and R. -H. Yue, 
``Reentrant phase transitions and triple points of topological AdS 
black holes in Born-Infeld-massive gravity,''
\href{https://doi.org/10.1155/2017/3819246}{\emph{Advances in High Energy Physics} (2017)
3819246}, [\eprint{1707.04101}].


\bibitem{Johnson} C. V. Johnson, ``Holographic heat engines,'' 
\href{https://iopscience.iop.org/article/10.1088/0264-9381/31/20/205002}
{\emph{Class. Quant. Grav.} 31 (2014) 205002}, [\eprint{1404.5982}].

\bibitem{Xu3} H. Xu, Y. Sun, L. Zhao, ``Black hole thermodynamics and heat engines 
in conformal gravity,''
\href{https://www.worldscientific.com/doi/abs/10.1142/S0218271817501516}
{\emph{Int. J. Mod. Phys. D} 26 (2017) no.13, 1750151}, [\eprint{1706.06442}].

\bibitem{Visser} M. R. Visser, ``Holographic thermodynamics requires a chemical 
potential for color,'' [\eprint{2101.04145}].

\bibitem{CKM} W. Cong, David Kubiznak, Robert B. Mann, 
``Thermodynamics of AdS black holes: central charge criticality,''
\href{https://journals.aps.org/prl/abstract/10.1103/PhysRevLett.127.091301}
{\emph{Phys. Rev. Lett.} 127, 091301 (2021)}, 
[\eprint{2105.02223}]

\bibitem{Alfaia} R. B. Alfaia, I. P. Lobo, L. C. T. Brito, 
``Central charge criticality of charged AdS black hole surrounded by different fluids,'' 
[\eprint{2109.06599}].

\bibitem{Kastor2} D. Kastor, S. Ray, J. Traschen, ``Chemical potential in the 
first law for holographic entanglement entropy,'' 
\href{https://link.springer.com/article/10.1007\%2FJHEP11\%282014\%29120}
{\emph{JHEP} (2014) 120}, [\eprint{1409.3521}].

\bibitem{Zhang} J. -L. Zhang, R. G. Cai, H. Yu, ``Phase transition and thermodynamical 
geometry of Reissner-Nordstr\"om-AdS black holes in extended phase space,''
\href{https://journals.aps.org/prd/abstract/10.1103/PhysRevD.91.044028}
{\emph{Phys. Rev. D} 91 (2015) 044028}, [\eprint{1502.01428}].

\bibitem{Karch} A. Karch, B. Robinson, ``Holographic black hole chemistry,''
\href{https://link.springer.com/article/10.1007\%2FJHEP12\%282015\%29073}
{\emph{JHEP} (2015) 1-15}, 
[\eprint{1510.02472}].

\bibitem{Maity} R. Maity, P. Roy, T. Sarkar, ``Black hole phase transitions and the 
chemical potential,'' 
\href{https://linkinghub.elsevier.com/retrieve/pii/S0370269316307407}
{\emph{Phys. Lett. B} 765 (2017) 386–394}, [\eprint{1512.05541}].

\bibitem{Wei} S. W. Wei, B. Liang, Y. X. Liu, ``Critical phenomena and chemical 
potential of a charged AdS black hole,'' 
\href{https://journals.aps.org/prd/abstract/10.1103/PhysRevD.96.124018}
{\emph{Phys. Rev. D} 96 (2017) 124018}, [\eprint{1705.08596}].

\bibitem{Rafiee:2021hyj}
M.~Rafiee, S.~A.~H.~Mansoori, S.~W.~Wei and R.~B.~Mann,
``Universal criticality of thermodynamic geometry for boundary conformal field theories in 
gauge/gravity duality,''
[\eprint{2107.08883}].

\bibitem{Tian1} \blue{Y. Tian, X.-N. Wu, H. Zhang, ``Holographic Entropy Production,'',
\href{https://link.springer.com/article/10.1007\%2FJHEP10\%282014\%29170}
{\emph{JHEP} 1410:170 (2014)},
[\eprint{1407.8273}].}

\bibitem{Tian2} \blue{Y. Tian, ``A topological charge of black holes,''
\href{https://iopscience.iop.org/article/10.1088/1361-6382/ab5343}
{\emph{Class. Quantum Grav.} 36 (2019) 245001},
[\eprint{1804.00249}].}

\bibitem{Zhao3} \blue{T. Wang, L. Zhao, ``Black hole thermodynamics is extensive 
with variable Newton constant,''
\href{https://doi.org/10.1016/j.physletb.2022.136935}
{\emph{Phys. Lett. B} 827 (2022) 136935}, 
[\eprint{2112.11236}]}.

\bibitem{Zhao4} \blue{L. Zhao, ``Thermodynamics for general rotating black holes 
with variable Newton constant,''  
\href{https://doi.org/10.1088/1674-1137/ac4f4c}{\emph{Chinese Phys. C}, (2022) in press}, 
[\eprint{2201.00521}]}.



\end{thebibliography}
\end{document}